\documentclass[preprint2]{aastex}

\usepackage{color,soul}
\usepackage{rotating}

\newcommand{\kms}{\,km\,s$^{-1}$} 
\newcommand\error{$\pm$}

\shorttitle{New halo stars of the GCs M3 and M13}
\shortauthors{Navin, Martell and Zucker}

\begin{document}

\title{New halo stars of the Galactic globular clusters M3 and M13 in the LAMOST DR1 Catalog}

\author{Colin~A.~Navin\altaffilmark{1}}
\affil{Department of Physics and Astronomy, Macquarie University, Sydney 2109, Australia.}
\email{colin.navin@mq.edu.au}
\author{Sarah~L.~Martell}
\affil{School of Physics, University of South Wales, Sydney 2052, Australia}
\email{s.martell@unsw.edu.au}
\and
\author{Daniel~B.~Zucker\altaffilmark{1}\altaffilmark{2}}
\affil{Department of Physics and Astronomy, Macquarie University, Sydney 2109, Australia.}
\email{daniel.zucker@mq.edu.au}
\altaffiltext{1}{Astronomy, Astrophysics and Astrophotonics Research Centre, Macquarie University, Sydney, NSW, 2109, Australia}
\altaffiltext{2}{Australian Astronomical Observatory, PO Box 915, North Ryde, NSW 1670, Australia.}

\begin{abstract}
M3 and M13 are Galactic globular clusters with previous reports of surrounding stellar halos. We present the results of a search for members and extratidal cluster halo stars within and outside of the tidal radius of these clusters in the LAMOST Data Release 1. We find seven candidate cluster members (inside the tidal radius) of both M3 and M13 respectively. In M3 we also identify eight candidate extratidal cluster halo stars at distances up to $\sim$9.8 times the tidal radius, and in M13 we identify 12 candidate extratidal cluster halo stars at distances up to $\sim$13.8 times the tidal radius. These results support previous indications that both M3 and M13 are surrounded by extended stellar halos, and we find that the GC destruction rates corresponding to the observed mass loss are generally significantly higher than theoretical studies predict. 
\end{abstract}


\section{Introduction}
\label{intro}

Globular clusters (GCs) lose stars through both internal processes such as stellar evolution and two-body relaxation, and external influences such as tidal disruption, dynamical friction and gravitational shocks due to passages close to the bulge and through the disks of their host galaxies. Consequently, at different epochs and even different points in their orbits, GCs may be less stable and have ex-member stars surrounding them or in associated tidal tail structures (e.g. \citealt{Vesperini:1997lr}, \citealt{Gnedin:1997lr}, \citealt{Baumgardt:2003qy}). The existence and properties of these extratidal stars can tell us how a GC has evolved since its formation via both internal dynamics and external influences of the host galaxy on the cluster (e.g. \citealt{Chernoff:1986fk}, \citealt{Vesperini:1997lr} and \citealt{Gnedin:1997lr}). It is necessary to understand these processes to understand the initial properties of the Galactic globular cluster system. They can also tell us about the host galaxy itself in several ways. The lost stars contribute to a galaxy's stellar population, and observed tidal tails may be used as tracers of the galactic gravitational potential (e.g. \citealt{Kupper:2015lr} and  \citealt{0004-637X-712-1-260}). They can also be an indicator of the formation history of the host galaxy, as some GCs are believed to be part of dwarf galaxies that are accreted (e.g.  \citealt{Mackey:2004uq}, \citealt{Bellazzini:2004fk}, \citealt{EAS:8463951}, \citealt{Carballo-Bello:2014lr}, \citealt{Marino:2014lr} and \citealt{Da-Costa:2015lr}).

\citet{Grillmair:1995lr} made the first discovery of Galactic GC tidal tails using automated star counts obtained from scanned UK Schmidt plates of 12 southern Galactic halo clusters. Likely cluster members on the main sequence, giant branch and horizontal branch were selected within empirically determined colour-magnitude envelopes.  Obvious tidal structures were visible in the two-dimensional surface density maps they constructed. Subsequent studies have found evidence for tidal tails and/or surrounding stellar halos in over 30 other GCs (e.g. \citealt{Leon:2000qy}, \citealt{1538-4357-641-1-L37}, \citealt{Kunder:2014lr}, \citealt{Navin:2015lr}, and \citealt{Anguiano:2016lr}) and even in the GCs of the Andromeda Galaxy (M31) \citep{Grillmair:1996fk}.

An indicator of the likelihood that GC stars will be found outside the tidal radius is the cluster destruction rate (or its inverse, the dissolution or destruction time). Significant simulations of the dynamical evolution of the Galactic globular cluster system include \citet{Aguilar:1988lr} and \citet{Hut:1992fk}. \citet{Gnedin:1997lr} calculated the total destruction rate of 119 GCs using simulations which included evaporation and disk and bulge gravitational shocks. They found that the present day destruction time was similar to the typical GC age. To estimate how many GCs have been destroyed since the formation of the Galaxy they favoured a scale-free power-law for the lifetime destruction rate. They concluded that the surviving population of GCs was a small fraction of those originally formed, and that a large fraction of the stars in the Galactic bulge and halo originated in GCs. A later study \citep{Dinescu:1999uq} of 38 GCs included proper motion data instead of statistically assigning velocities. They  concluded that the orbits used in \citet{Gnedin:1997lr} were more destructive than are actually observed, so destruction rates for many clusters may have been overestimated. \citet{Mackey:2004uq} estimated that $\sim$100 of the present Milky Way GC population were formed in the Galaxy. Given that those with reasonably concentrated core radii ($r_c < 2$ pc) are less likely to be disrupted, they calculated that at least 50 percent of clusters have been destroyed over the last Hubble time. \citet{Mackey:2005qy} estimated that the present population is 67 percent of the original, using observational differences in properties of 'young halo', 'old halo' and 'bulge/disc' Galactic GC subsystems. Detailed simulations by \citet{Moreno:2014qy} calculated orbits, tidal radii and destruction rates due to bulge-bar and disk shocking for a sample of 63 Galactic GCs using six-dimensional data in axisymmetric and non-axisymmetric Galactic potentials including a Galactic bar and a 3D model for the spiral arms. 

It is likely that a significant fraction of stars in the bulge and halo of the Milky Way (MW) originated in GCs. \citet{Martell:2010fk} studied the SDSS-II/SEGUE spectra of $\sim$1900 G and K-type halo giants and found that 2.5\% showed abundance patterns only previously found in GC stars. They inferred that up to  50\% of halo field stars initially formed within GCs. A further study \citep{Martell:2011qy} of 561 low metallicity halo giant stars in SDSS-II/SEGUE 2 concluded, based on prevailing models of GC formation at the time, that a minimum of 17\% of the present-day mass of the stellar halo originally formed in GCs.

GCs tidal debris also acts as indicators of a host galaxy's gravitational potential as the extratidal stars spread out in a stream that traces the orbit of its progenitor. Recent work on this includes \citet{Kupper:2015lr}, who used the stellar stream associated with Pal 5 to constrain the Galactic mass within its apogalactic radius. \citet{0004-637X-712-1-260} used the long narrow GD-1 stream of stars, likely to be from a defunct tidally disrupted GC, to constrain the circular velocity at the Sun's radius and the Galactic total potential flattening.

\section{Data and initial selection process}
\label{data}

There has already been a study searching for open and GC members \citep{1674-4527-15-8-1197} in the LAMOST spectroscopic survey \citep{Zhao:2006kx}. The survey footprint covers a number of Northern hemisphere GCs, and therefore had potential as a dataset to search for GC extratidal stars that have been observed in the program.  

The LAMOST survey is a low/medium resolution spectroscopic survey of the Northern hemisphere which aims to obtain the spectra of 10,000,000 objects, including stars, galaxies and quasars. It uses the Guoshoujing Schmidt telescope located at the Xinglong Observatory in China. This telescope has a clear aperture of 4.0 m and a 5\degr\ field-of-view, with 4,000 optical fibres of 3 arcsec diameter leading to 16 spectrographs. The spectral range is 3650--9000 \AA\, with a limiting magnitude r = 20 at a resolution of R = 500.

Data Release 1 (DR1) \citep{Luo:2015lr} contains data from the Pilot and First Year Surveys and is now publicly available for the general astronomical community. It contains 1,944,329 stellar spectra, with the DR1 AFGK Stars Catalog containing 1,061,918 high quality spectra (later references in this paper to DR1 or the DR1 Catalog refer specifically to the DR1 AFGK Stars Catalog). As well as basic data such as positions and magnitudes from the input catalogs, these stars have values for stellar atmospheric parameters (effective temperature ($T_{eff}$), surface gravity (log$g$) and [Fe/H]) and heliocentric radial velocity ($V_r$) derived using the LAMOST Stellar Parameter Pipeline (LASP).

Differentiating stellar cluster members from field stars is possible because cluster members are expected to share characteristics derived from their common origin. It is possible to identify candidate cluster members as a clump of stars in the parameter space defined by position, $V_r$, stellar atmospheric parameters ($T_{eff}$, log$g$ and [Fe/H]), photometry and proper motions. The gold standard for identification is detailed abundance matching from high resolution spectroscopy, but large samples of candidates need to be cleaned in order to produce likely targets for this method. 

Our first step was to identify target GCs that might have easily identifiable members or extratidal stars in the DR1 Catalog. We selected GCs within the survey area of LAMOST (declination $-$10\degr\ to +90\degr) that had relatively high heliocentric radial velocities ($|V_r| >100$ \kms) to simplify differentiation of candidate stars from field stars in the same area of sky. There are 19 northern hemisphere GCs that satisfy these criteria. For each of these GCs we selected stars from the DR1 Catalog that were within a radius of 5\degr\ of the GC central position to encompass both a wide area within and outside the tidal radius, and that had a $V_r$ within \error20~\kms\ of the GC $V_r$. 

We found candidate stars in the DR1 Catalog around ten GCs that satisfied these criteria: NGC~4147, M3, NGC~5466, M13, NGC~6229, M92 (NGC~6341), NGC~6426, NGC~6535, NGC~7078 and Pal~2. We discarded seven clusters as potential targets as they either had very few candidate stars or their $V_r$ did not sufficiently differentiate them from the bulk of surrounding field stars. The three remaining GCs (M3, M13 and NGC~6229) had both a significant number of potential candidate stars and a $V_r$ that was significantly distinct from field stars in the sky area.

Finally, we also rejected NGC~6229 as a potential target as the DR1 Catalog stars within a 5\degr\ radius of the GC central position are brighter than the cluster RGB. LAMOST exposes bright, medium and faint plates that result in cuts at r = 14, 16 and 19 mag, respectively. All the DR1 Catalog stars in this sky area that satisfied the $V_r$ criteria are brighter than V = 16 but on the $V$ versus $V-K$ CMD the cluster RGB (defined by a PARSEC isochrone; \citealt{Bressan:2012fk}) is mostly fainter than V = 16. It is likely that no faint fields were taken in this sky area during observations for DR1, hence there are no likely members of NGC~6229 in the DR1 Catalog.  Although we found a number of potential candidate stars based on $V_r$, because we were not solely relying on radial velocity for selection we excluded those stars which were not consistent with the cluster RGB.

We investigate the two remaining clusters, M3 and M13 in this study.  From initial samples of candidate stars for M3 and M13 selected by $V_r$ and position (Sect. \ref{vr}), we utilise photometry, log$g$ and log($T_{eff}$) (Sect. \ref{psp}), proper motions (Sect. \ref{pm}) and metallicities (Sect. \ref{metal}) to clean our samples. We look at the spatial distribution of the candidate member samples with respect to the adopted tidal radii in Sect. \ref{sd} and present and discuss the final list of candidate cluster members and cluster halo stars in Sect. \ref{discussion}.

\section{M3 and M13 background}
\label{M3_and_M13}

\begin{deluxetable}{l c c c c c c c c c c c c}
\tabletypesize{\footnotesize}
\setlength\tabcolsep{3pt}
\tablewidth{0pt}
\tablecaption{Basic data on M3 and M13\label{GC_data_table}}
\tablehead{
\colhead{}   & \colhead{$V_t$} & \colhead{$R_{\sun}$} & \colhead{$R_{gc}$} &
\colhead{[Fe/H]} & \colhead{$V_r$} & \colhead{$\mu_{\alpha}cos(\delta)$} & \colhead{$\mu_{\delta}$} & 
\colhead{$c$} & \colhead{$r_c$} & \colhead{$r_t$}   \\
\colhead{}   & \colhead{(mag)}   & \colhead{(kpc)}   & \colhead{(kpc)}  &
\colhead{}   & \colhead{(\kms)}  & \colhead{(mas~yr$^{-1}$)}  & \colhead{(mas~yr$^{-1}$)}    & 
\colhead{(arcmin)}  & \colhead{(arcmin)}  & \colhead{(arcmin)}
}
\startdata
M3 (NGC~5272)    &  6.19  & 10.2  & 12.0  &  $-$1.53  &  $-$147.6   &  $-$1.1  &  $-$2.3  &  1.89  &  0.37  &  28.7 \\
M13 (NGC~6205)  &  5.78  &   7.1  &   8.4  &  $-$1.50  &  $-$244.2   &  $-$0.9  &   5.5  &  1.53  &  0.62  &  21.0 \\
\enddata
\tablecomments{Data from \citet{Harris:1996lr}\ (2010 edition) catalog. $V_t$ is the integrated V magnitude of the cluster. Proper motions $\mu_{\alpha}cos(\delta)$ and $\mu_{\delta}$ from \citet{Dinescu:1999uq}. Central concentration ($c$) and core radius ($r_c$) from \citet{McLaughlin:2005qy}, tidal radius ($r_t$) calculated from ${c} = {log{\frac{r_t}{r_c}}}$}
\end{deluxetable}

Both M3 and M13 are extremely well-studied observationally (e.g. both are in the SDSS footprint) and theoretically, and both have a complicated history regarding the possibility of extratidal stars. Searches (see below) for the existence of surrounding stellar halos outside the tidal radius for M3 have only sometimes been successful; observationally the case for extratidal stars around M13 is stronger. The search of \citet{1674-4527-15-8-1197} for both open and globular cluster members in the LAMOST Data Release 2 (DR2) found M3 and M13 as the only two GCs with identified cluster members. Basic data for both clusters are presented in Table~\ref{GC_data_table}.

M3 was one of the GCs studied in a search for tidal tails by \citet{Leon:2000qy}. Member stars on Schmidt plates were identified using CMDs, and a star-count analysis was performed on the selected cluster stars. All 20 clusters analysed showed evidence of tidal tails with projected directions preferentially towards the Galactic centre.  M3 showed evidence of an extension perpendicular to the Galactic plane, with the caveat that features in the cluster's density contours may have been introduced by contamination of the stellar sample. The most important mass loss process, based on the shape of the tidal tail as well as the GC's position, orbit and proper motion, was disk-shocking. They noted that the importance of the interaction of GCs and the Galaxy was underlined by the detection of tidally stripped stars.  However, later studies have found ambiguous evidence for the existence of extratidal stars around M3. \citet{1538-4357-639-1-L17} used optimally filtered star counts for SDSS photometry, but found no evidence for tidal streams. \citet{Jordi:2010lr} also used SDSS photometry with a colour-magnitude weighted counting algorithm to search for extratidal features. They controlled carefully for sample contamination by background galaxies and quasars as well as dust extinction. The number density profile for M3 did not show any change in slope hinting at an extratidal halo, and density contours did not reveal any large scale tidal structure. They noted that this result was consistent with the \citet{Gnedin:1997lr} simulations. However, \citet{Chen:2010qy} used star count analysis of the brighter members in the outer part of 116 Galactic GCs in the 2MASS \citep{2006AJ....131.1163S} Point Source Catalog. They listed M3 as one of 31 GCs that exhibited outlying filamentary or clumpy density enhancements, indicating the presence of stripped stellar members. This clumpiness was not, however, consistent with that found by \citet{Leon:2000qy}, and they did note that some clumpiness could arise from statistical fluctuations. \citet{Carballo-Bello:2014lr} used both cross-correlation and isochrone fitting methods on wide-field photometric data from a variety of sources on a sample of 23 GCs. They were searching for evidence of an accretion origin for the GCs in the form of residual debris around the clusters from their now defunct host dwarf galaxies. They reported no detections of any surrounding extratidal stellar structure around M3 and 13 other GCs, indicating that these GCs were not accreted along with a host dwarf galaxy. They noted that an early accretion time or low mass of an original host dwarf galaxy would also reduce the likelihood of the presence of extratidal stars and so did not rule out this scenario entirely.

M3 has a perigalacticon of 5.5$\pm$0.8 kpc and is 12.0 kpc from the Galactic centre, quite close to its apogalacticon of 13.7$\pm$0.8 kpc \citep{Dinescu:1999uq}. In theoretical studies, the previously cited \citet{Gnedin:1997lr} paper calculated total destruction rates for M3. For an isotropic GC system distribution with a Weinberg adiabatic conservation factor for the shock processes, they quote $3.98\times10^{-12}$\ and\ $5.42\times10^{-12}~yr^{-1}$\ for two different Galactic models. This was $\sim$2 orders of magnitude less than the median destruction rates for their sample of 119 GCs for the same setup. \citet{Moreno:2014qy} calculated the total destruction rate due to bulge-bar and disk shocking for M3. Depending on the models used this ranged from $1.62\times10^{-12}$\ to\ $2.97\times10^{-12}~yr^{-1}$, somewhat lower than their median values. Both these studies indicate that M3 is relatively stable, with expected low rates of star loss.

The case for M13 is somewhat stronger observationally, with a number of searches finding evidence for extratidal stars. \citet{Lehmann:1997lr} used automated star counts on scanned Schmidt plates to obtain projected density profiles and calculate structural parameters such as the tidal radii of seven GCs. They found that M13 (and other GCs) showed evidence of tidal tails in the form of increased surface density outside the calculated tidal radius of 23.8 arcmin. The previously cited \citet{Leon:2000qy} study also included M13. They were not able to search over a very large sky area, but inside their adopted tidal radius of 56 pc they noted an extension towards the Galactic centre in plots of density contours. Bulge shocking was considered to be the most likely process that would cause any loss of stars. \citet{Jordi:2010lr} also detected a halo of extratidal stars extended slightly in the direction of motion in a similar position as the disturbed contours noted in \citet{Leon:2000qy}. The previously cited \citet{Chen:2010qy} study found M13 also showed possible outlying clumpy stellar debris. As with M3, this structure did not correspond with that found by the \citet{Leon:2000qy} study.

M13 has an apogalacticon of 21.5$\pm$4.7 kpc and is currently located reasonably close to its perigalacticon of 5.0$\pm$0.5 kpc \citep{Dinescu:1999uq} at a distance of 8.4 kpc from the Galactic centre. \citet{Gnedin:1997lr} quoted total destruction rates for M13 of $1.47\times10^{-11}$\ and\ $1.02\times10^{-11}\ yr^{-1}$\ for their two different galactic models. These are more than an order of magnitude less than the sample median destruction rate. The destruction rates calculated by \citet{Moreno:2014qy} for M13 were, however, much lower at  $6.20\times10^{-14}$\ to\ $1.04\times10^{-13}~yr^{-1}$.  As with M3, these theoretical studies show that M13 is expected to be relatively stable.

\section{Heliocentric radial velocities}
\label{vr}

\begin{figure}
\plotone{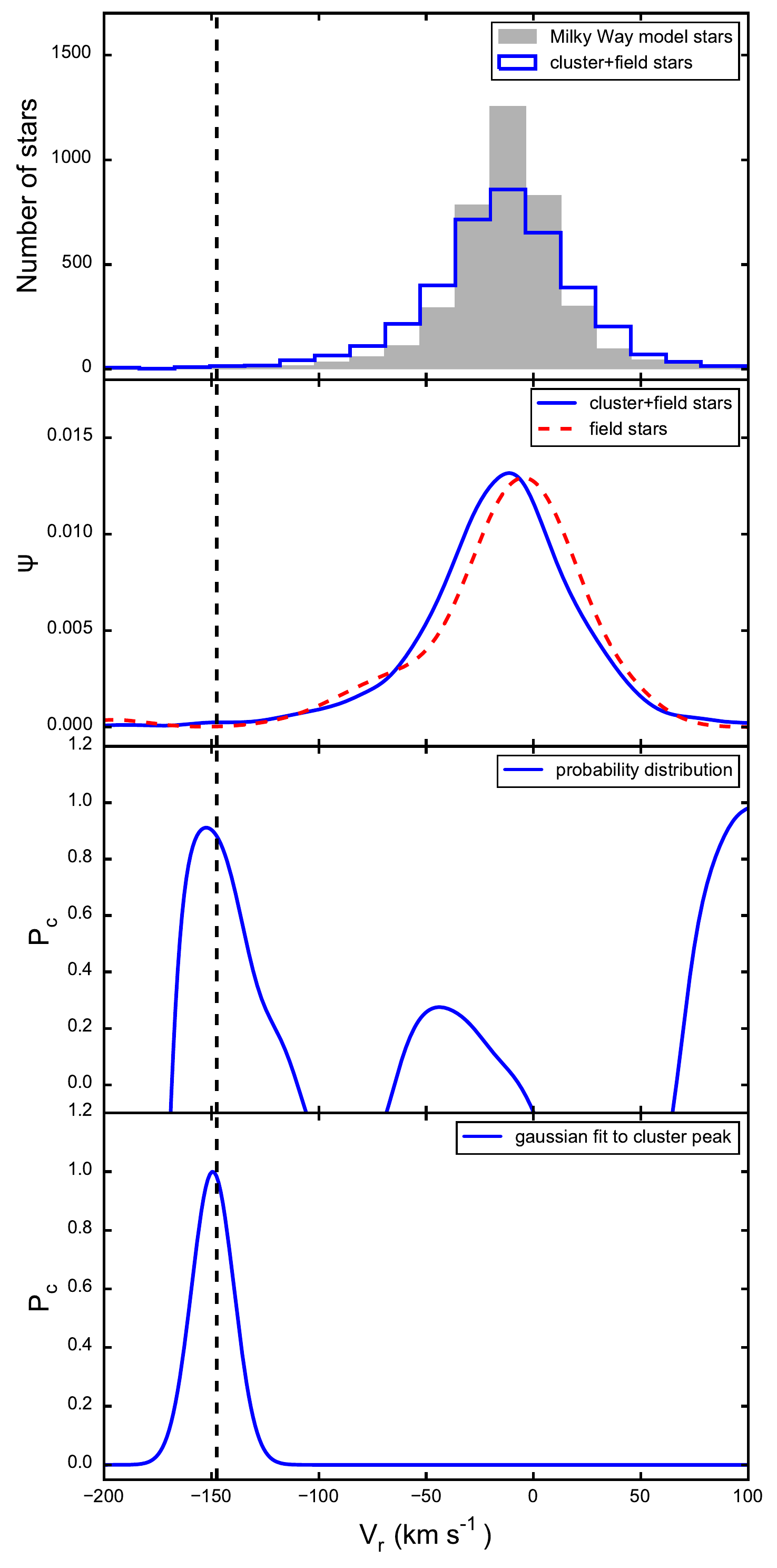}
\caption{First panel: histogram of all the observed stars within a radius of 5\degr\ of M3's central position with the predicted $V_r$ distribution of the Milky Way model generated by the Galaxia code \citep{Sharma:2011lr} overplotted as a grey histogram. Second panel: The solid blue line and the dashed red line show the Gaussian KDEs of the $V_r$ of the observed stars and the field stars, respectively. Third panel: The solid blue line shows the $V_r$ membership probability distribution. Fourth panel: The solid blue line shows the Gaussian fit to the cluster peak. The vertical dashed line shows the GC $V_r$.}
\label{NGC5272_M3_Vr_hist_pdfs}
\end{figure}

\begin{figure}
\plotone{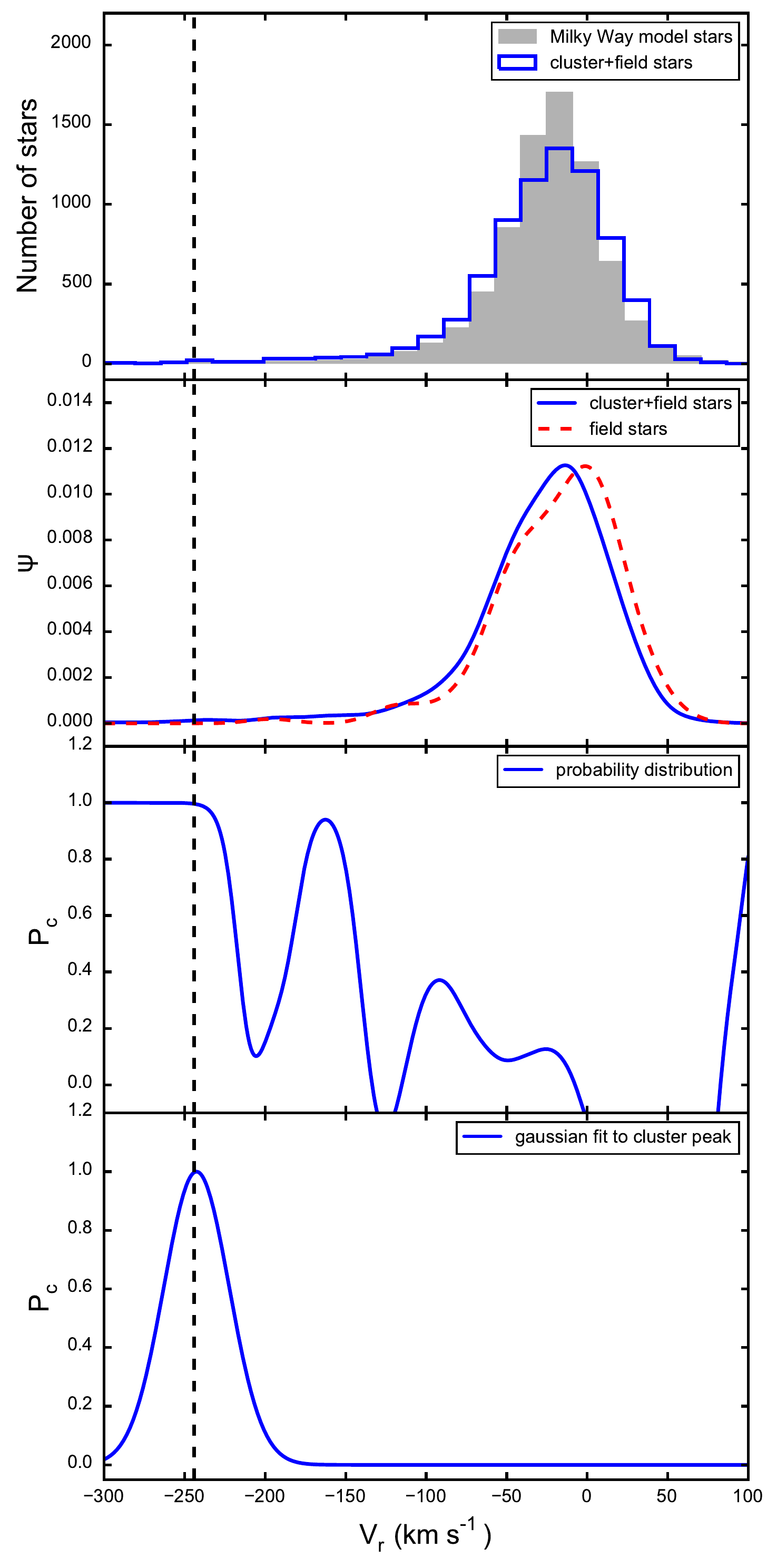}
\caption{Same as Figure \ref{NGC5272_M3_Vr_hist_pdfs} for M13}
\label{NGC6205_M13_Vr_hist_pdfs}
\end{figure}

The first panels of Figure \ref{NGC5272_M3_Vr_hist_pdfs} and Figure \ref{NGC6205_M13_Vr_hist_pdfs} show $V_r$ histograms of all the observed stars, with the predicted $V_r$ distribution of the Milky Way model (see below) generated by the Galaxia code \citep{Sharma:2011lr}, overplotted as a grey histogram. 

There were 4426 entries within a radius of 5\degr\ of the central position of M3 in the DR1 Catalog; 547 stars ($\sim$14\%) were duplicates (observed multiple times) leaving 3879 unique stars. For M13 there were 8265 entries within a radius of 5\degr\ of the central position; 910 stars ($\sim$12\%) were duplicates leaving 7355 unique stars. These numbers are consistent with the DR1 Catalog in general where $\sim$18\% of the targets in total have been observed more than one time \citep{Luo:2015lr}.

For the duplicate stars we calculated $V_r$ (and also $T_{eff}$, log$g$ and [Fe/H] below) from the DR1 values by taking a mean weighted by the respective errors e.g.:
\begin{equation}
\overline{V_r} = \frac{\sum{V_i/{\delta_i}^2}}{{\sum{1/{\delta_i}^2}}}
\end{equation}
where  $V_i$ is the \textit{i}th $V_{r}$ and $\delta_i$ is the corresponding \textit{i}th error.

We calculated a simple quadrature error in the parameter by e.g.:
 \begin{equation} \overline {\Delta V_r} = \frac{\sqrt{\sum{\delta_i}^2}}{\sqrt{N}} 
 \end{equation} where N is the number of observations of the star.
 
We used the Galaxia code to generate a synthetic catalogue of stars to compare with the observations. Galaxia predicts radial velocities, metallicities and other properties of a stellar sample given colour, magnitude and spatial constraints. We used the Besan\c con Milky Way model \citep{Robin:2003fk} for the disk and the simulated N-body models of \citet{Bullock:2005fk} for the stellar halo. We generated 10 models covering a 5\degr\ radius around the cluster centre position with no colour or magnitude restrictions. The LAMOST observational target catalog selection strategy \citep{Luo:2015lr} is complex and, of course, for many practical reasons an input catalogue may differ from the eventual products of a survey. Therefore, pragmatically, we used a simple magnitude cut ($9.0 \leq J \leq 12.5$) to construct the final $V_r$ distribution. Finally, we summed and normalised the $V_r$ distributions to have the same number of stars as the number of observed stars.

The sample of 3879 unique stars for M3 has a mean and standard deviation of the $V_r$ distribution of $\overline{V_r}\sim-$16$\pm$0.6 \kms\ and $\sigma_{V_r}\sim\ $40$\pm$0.5 \kms\ respectively. In comparison the average Milky Way model distribution has $\overline{V_r}\sim-$13$\pm$0.5 \kms\ and $\sigma_{V_r}\sim\ $30$\pm$0.3 \kms. For the 7355 unique stars in the M13 sample, $\overline{V_r}\sim-$29$\pm$0.6 \kms\ and $\sigma_{V_r}\sim\ $48$\pm$0.4 \kms; the corresponding figures for the model are $\overline{V_r}\sim-$25$\pm$0.4 \kms\ and $\sigma_{V_r}\sim\ $38$\pm$0.3 \kms. Both distributions show a prominent peak near the mean $V_r$ that we identify as predominantly Galactic disc or halo field stars (non cluster members) plus an (unknown) number of cluster members.
 
We applied offsets to the $V_r$ values in the DR1 Catalog to correct for systematic errors between the DR1 Catalog and the MWSC catalog \citep{Kharchenko:2013fk} noted in \citet{1674-4527-15-8-1197}. For M3 the offsets are in the sense: DR1 $V_r\ -$ \citet{Harris:1996lr} $V_r = -$6.4  \kms\ and for M13 the offsets are in the sense: DR1 $V_r\ -$ \citet{Harris:1996lr} $V_r$ = $-$6.9 \kms.

We calculated the cluster membership probabilities $P_c$ of stars on the basis of $V_r$ using the method of \citet{Frinchaboy:2008qy}. We assumed that stars between two and three times the $r_t$ of the cluster central position produced a sample of non-member field stars in the sky area of the cluster. There are 136 and 147 stars in these field samples for M3 and M13 respectively. However, as we were looking for extratidal stars that might well be in this sky area, it is possible to improve the field sample by removing stars that we already suspect are associated with the cluster. For M3 there is one star between two and three times the $r_t$  with $V_r$ within \error20~\kms~of the GC $V_r$, so we removed that star from the field sample. This left 135 stars in the field samples for M3.

First we made smoothed kernel density estimates (KDEs), i.e. the histograms were convolved with Gaussians with widths set by the measurement errors of (i) $\psi_{c+f}$ - the complete $V_r$ distribution (i.e. cluster stars plus field stars out to a radius of 5\degr\ from the cluster centre), and (ii)  $\psi_{f}$ - the $V_r$ distribution of field stars between two and three times the $r_t$. This is shown in the second panels on Figure \ref{NGC5272_M3_Vr_hist_pdfs} and Figure \ref{NGC6205_M13_Vr_hist_pdfs}. The formula for membership probability used by \citet{Frinchaboy:2008qy} is:
\begin{equation}
P_c = \frac{\psi_{c+f} - \psi_f}{\psi_{c+f}}
\end{equation}
The membership probability distributions are shown in the third panels on Figure \ref{NGC5272_M3_Vr_hist_pdfs} and Figure \ref{NGC6205_M13_Vr_hist_pdfs}. 

We then applied Gaussian fits to the distributions around the cluster peaks in $V_r$ (fourth panel on Figure \ref{NGC5272_M3_Vr_hist_pdfs} and Figure \ref{NGC6205_M13_Vr_hist_pdfs}). The KDE of field stars was extremely small in the region of both cluster peaks so the membership probability was very sensitive to that value. This resulted in probability values slightly in excess of one near the peak of the distribution, so we normalised the distributions so that the maximum values at the peaks were one. We then used the standard deviations ($\sigma$) of these Gaussian fits to select candidate cluster members that were within \error$2~\sigma$ of the GC $V_r$ and to determine their membership probabilities. 

The mean and standard deviation of the Gaussian fit to the cluster peak for M3 for $V_r$ are $-$152.7 and 15.4~\kms~respectively; the literature $V_r$ is $-$147.6 \error\ 0.2~\kms. For M13 the mean and standard deviation are $-$242.8 and 20.0~\kms~respectively; the literature $V_r$ is $-$244.2 \error\ 0.2~\kms. Members and extratidal halo stars of a GC are expected to share its $V_r$ signature, so for initial samples of candidate cluster members on the basis of $V_r$  we selected stars that were within \error$2~\sigma$ of the GC $V_r$. 

For M3 there are 57 stars in the DR1 Catalog that are within \error$2\sigma$ of the GC $V_r$ and inside a radius of 5\degr\ of the GC central position. Seven of these stars are duplicates (each observed twice) and so we calculated mean parameter values and errors for these stars as above. This left a sample of 50 unique stars that we refer to as our M3 candidate stars. The mean and standard deviation of the $V_r$ distribution of the candidate stars are $-$137.6 and 15.4~\kms. The $V_r$ of these stars suggests that they are possibly cluster members or cluster halo stars. Our Milky Way model predicts $\sim$18 stars in the same area of sky within \error$2~\sigma$ of the GC $V_r$ that have no relation to the cluster and would therefore result in false positives, so we also used other parameters to further clean the sample. 

There are 67 stars  with six duplicates within \error$2~\sigma$ of the $V_r$ and within 5\degr\ of the GC central position for M13. The mean and standard deviation of the $V_r$ distribution of the 61 unique stars left as our M13 candidate stars are $-$236.8 and 20.0~\kms\ respectively. As with M3, their $V_r$ suggests that these stars are possibly cluster members or cluster halo stars; for M13 our Milky Way model predicts a sample contamination of $\sim$28 stars.

Many DR1 catalog $V_r$ errors are fairly large for the stars selected by $V_r$ for both clusters. Typical errors in $V_r$ are quoted as 5~\kms\ \citep{1674-4527-15-8-1197}, however the mean $V_r$ error for our candidate members for M3 is 15.2~\kms\ and for M13 it is 17.1~\kms, similar to the standard deviation of the $V_r$ distribution of the candidate stars. Therefore it is likely that the errors of the LAMOST radial velocities are the largest contributor to the dispersion in the Gaussian fits to the cluster peaks.

\section{Photometry and stellar parameters (surface gravity and effective temperature)}
\label{psp}

\begin{figure*}
\includegraphics[width=168mm]{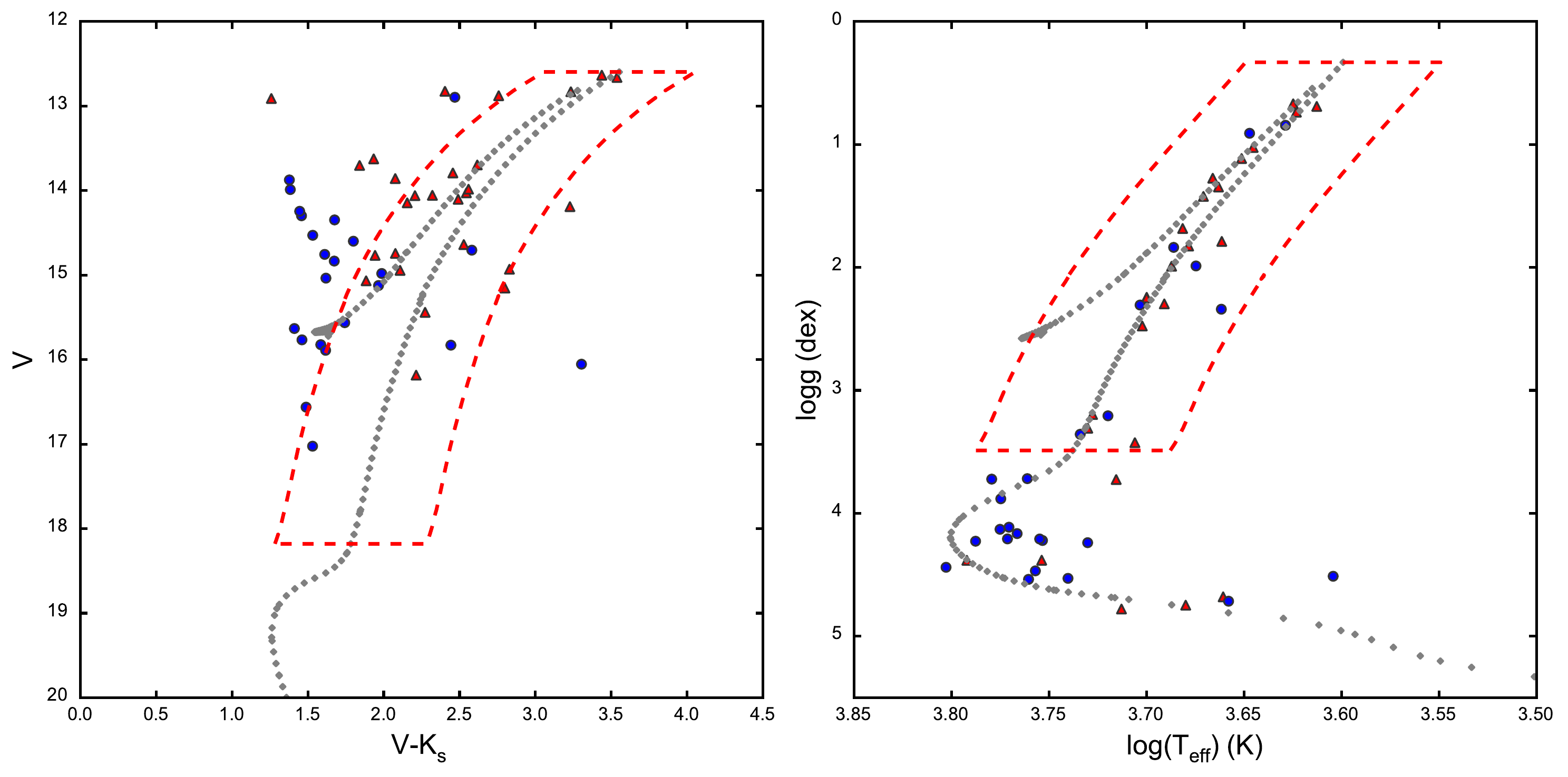}
\caption{Left-hand panel: $V$ versus $V-K_s$ CMD of candidate stars around M3. Right-hand panel: log($T_{eff}$) versus log$g$ diagram of candidate stars. The grey diamonds show the PARSEC isochrone \citep{Bressan:2012fk} for the GC. In the left panel the dashed red line shows the boundary for candidate selection based on $V$ versus $V-K_s$ photometry. The stars are colour-coded with red triangles as the  stars that are inside the $T_{eff}$ versus log$g$ limits box (the red dashed box on the right panel) and blue circles as stars outside the $T_{eff}$ versus log$g$ limits box. In the right panel the dashed red line shows the boundary for candidate selection based on $T_{eff}$ versus log$g$. The stars are colour-coded with red triangles as stars that are inside the $V$ versus $V-K_s$ limits box (the red dashed box on the left panel) and blue circles as stars that are outside the $V$ versus $V-K_s$ limits box.}
\label{NGC5272_M3_CMD_V_V_K_Teff_logg}
\end{figure*}

\begin{figure*}
\includegraphics[width=168mm]{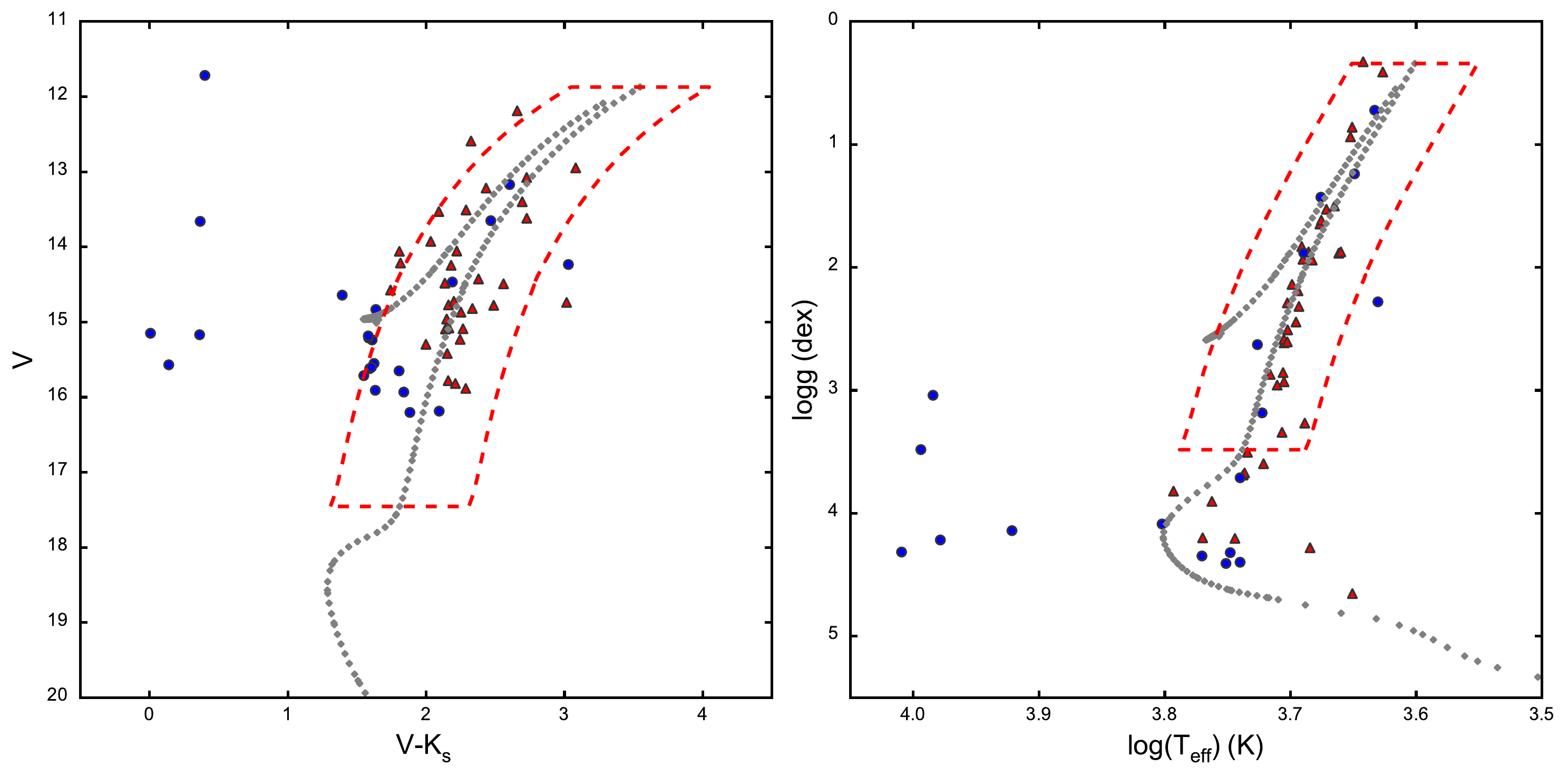}
\caption{Same as Figure \ref{NGC5272_M3_CMD_V_V_K_Teff_logg} for M13}
\label{NGC6205_M13_CMD_V_V_K_Teff_logg}
\end{figure*}

Photometric data were obtained from the UCAC4 catalog \citep{Zacharias:2013lr} containing V magnitudes from APASS \citep{Henden:2009qy} and $K_s$ magnitudes from 2MASS \citep{2006AJ....131.1163S}. Figure \ref{NGC5272_M3_CMD_V_V_K_Teff_logg} shows the $V$ versus $V-K_s$ CMD and the log($T_{eff}$) versus log$g$ diagrams for the 50 unique M3 candidate stars. The plots also show the PARSEC isochrone \citep{Bressan:2012fk} generated for M3. Input parameters for producing the isochrone are: age 12.6 Gyr from the Milky Way Star Clusters (MWSC) catalog \citep{Kharchenko:2013fk}; [Fe/H] = $-$1.50; [$\alpha$/Fe] = 0.3 \citep{Marin-Franch:2009fk} $\Rightarrow$ Z = 0.0079; E(B$-$V) = 0.01 $\Rightarrow A_v$ = 0.032; apparent visual distance modulus ($m_v - M_v$) = 15.07. The V magnitude of the horizontal branch $V_{HB}$ is 15.64 and this agrees with its position on the $V$ versus $V-K$ CMD.

The $V$ versus $V-K_s$ CMD and the log($T_{eff}$) versus log$g$ diagrams for the 61 unique M13 candidate stars are shown in Figure \ref{NGC6205_M13_CMD_V_V_K_Teff_logg}. Input parameters for generating the PARSEC isochrone are: age 12.6 Gyr from the MWSC catalog \citep{Kharchenko:2013fk}; [Fe/H] = $-$1.53; [$\alpha$/Fe] = 0.3 \citep{Marin-Franch:2009fk} $\Rightarrow$ Z = 0.0073; E(B$-$V) = 0.02 $\Rightarrow A_v$ = 0.064; apparent visual distance modulus ($m_v - M_v$) = 14.33.  $V_{HB}$ is 14.90 and this agrees with its position on the $V$ versus $V-K$ CMD. 

For both clusters, the magnitudes of  observed stars and the positions of the isochrones on the CMDs show that any GC members must be giants and that any dwarfs observed are foreground stars rather than cluster members. Therefore our $V$ versus $V-K$ and log($T_{eff}$) versus log$g$ limits were based on selecting stars close to the RGB.

The dashed red lines on the left panel show the $V$ versus $V-K_s$ boundaries we adopted for candidate selection with respect to the isochrone for M3. The boundaries are $\pm0.5$ mag in $V-K_s$ and $12.6 < V < 18.2$ mag to incorporate the $V$  magnitude range of the RGB. The dashed red lines on the right panel show the log($T_{eff}$) versus log$g$ boundaries we adopted. This incorporates a cut at log$g$ = 3.5 to separate dwarfs from giants and the width of the box is $\pm0.05$ in log($T_{eff}$). Altogether we removed 31 stars that were outside either of these boundaries. 

For M13 we adopted boundaries for the photometric selection of $\pm0.5$ mag in $V-K_s$ and $11.9 < V < 17.5$ mag and for the log($T_{eff}$) versus log$g$ boundaries we adopted the same cut at log$g$ = 3.5 and again the width of the box is $\pm0.05$ in log($T_{eff}$). We removed 29 stars from the list of candidates based on these boundaries.

\section{Proper motions}
\label{pm}

Observed proper motions can also be used to clean a sample of candidate stars if their proper motions do not match that of the cluster. For our purposes we adopted a limit of 10~mas~yr$^{-1}$ of the GC proper motion. However because it is difficult to make proper motion measurements in crowded stellar fields, we discounted proper motion as a necessary condition for membership for the candidate stars inside the tidal radius.

\begin{figure}
\plotone{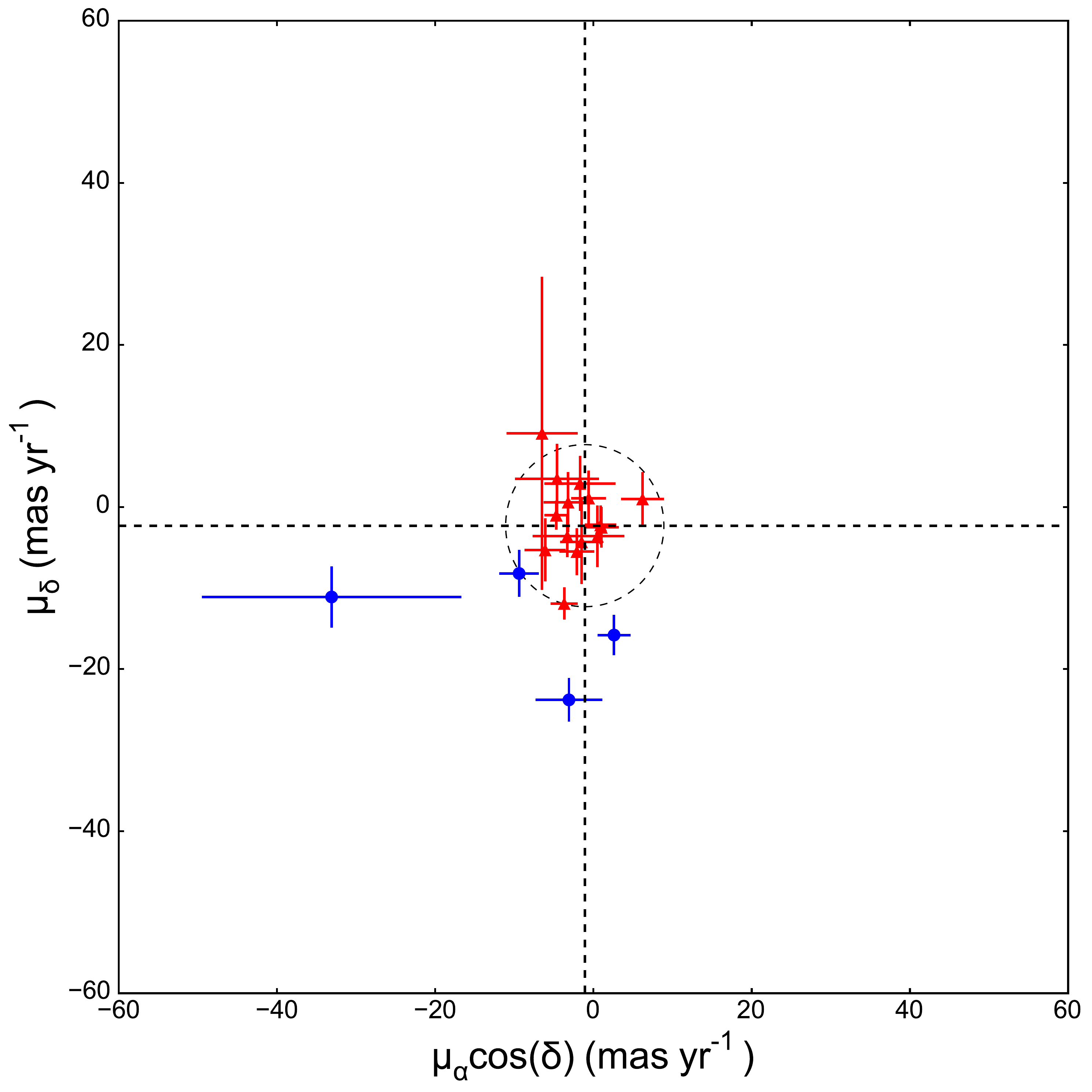}
\caption{Absolute proper motions of the candidate stars of M3. Red triangles denote stars outside the tidal radius with proper motions within 10~mas~yr$^{-1}$ of the GC proper motion, or stars that are inside the tidal radius. Blue circles denote stars outside the tidal radius that have proper motions more than 10~mas~yr$^{-1}$ different from the GC proper motion. The vertical and horizontal black dashed lines indicate the GC proper motion and the black dashed circle indicates the 10~mas~yr$^{-1}$ limit.}
\label{NGC5272_M3_proper_motions}
\end{figure}

\begin{figure}
\plotone{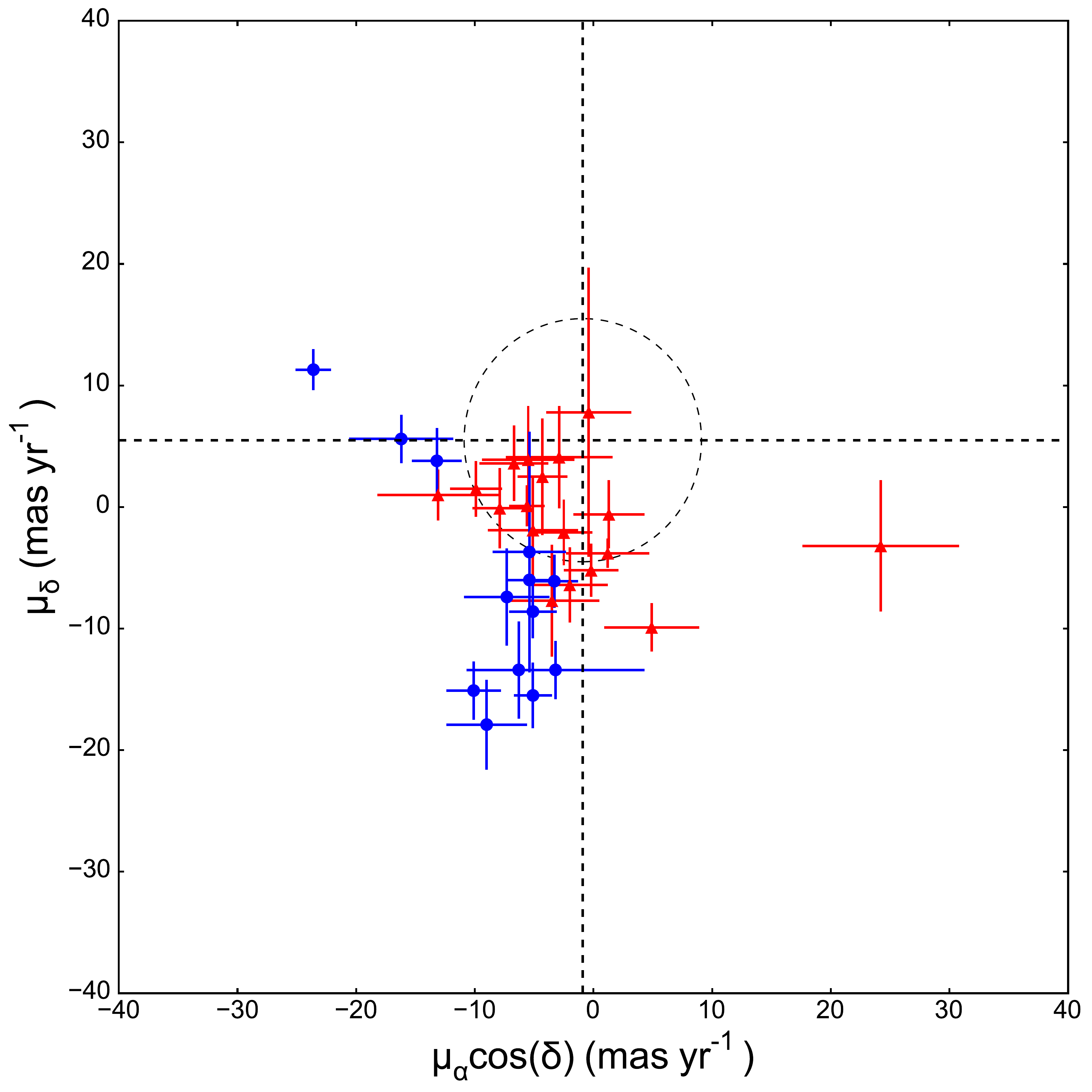}
\caption{Same as Figure \ref{NGC5272_M3_proper_motions} for M13.}
\label{NGC6205_M13_proper_motions}
\end{figure}

We obtained the absolute proper motions of all 19 remaining candidate stars of M3 from the UCAC4 catalog \citep{Zacharias:2013lr} and the GC proper motion ($\mu_{\alpha}\cos(\delta)\ = -1.1$\error0.51~mas~yr$^{-1}$, $\mu_{\delta}\ = -2.3$\error0.54~mas~yr$^{-1}$) from \citet{Dinescu:1999uq}. Figure \ref{NGC5272_M3_proper_motions} shows a plot of the absolute proper motions of the candidate stars. For stars outside the tidal radius, we accepted stars with proper motions that are within 10~mas~yr$^{-1}$ of the GC proper motion (black dashed circle on Figure \ref{NGC5272_M3_proper_motions}). Four of the 19 candidate stars that are outside the GC tidal radius had proper motions more than 10~mas~yr$^{-1}$ different from the GC proper motion so we eliminated these as candidates. We did not discount any stars that are inside the tidal radius as potential members on the basis of their proper motions. There were seven stars inside the tidal radius and we kept all these stars as candidates. 

UCAC4 has absolute proper motions for 31 of the 32 candidate stars of M13. The GC proper motion of the GC is $\mu_{\alpha}\cos(\delta)\ = -0.9$\error0.71~mas~yr$^{-1}$, $\mu_{\delta}\ = 5.5$\error1.12~mas~yr$^{-1}$ \citep{Dinescu:1999uq}. Figure \ref{NGC6205_M13_proper_motions} shows a plot of the absolute proper motions of the candidate stars. We eliminated 13 of the 25 stars that are outside the GC tidal radius with proper motions more than 10~mas~yr$^{-1}$ different from M13's proper motion. There were seven stars inside the tidal radius and we kept all these stars as candidates. We also retained in our list of candidates one star without proper motion data (obsid 50501031).

\section{Metallicities}
\label{metal}

We applied offsets to the [Fe/H] values in the DR1 Catalog to correct for systematic errors between the DR1 Catalog and the MWSC catalog \citep{Kharchenko:2013fk} noted in \citet{1674-4527-15-8-1197}, and the offsets in cluster [Fe/H] between the \citet{Harris:1996lr} catalog and the MWSC catalog \citep{Kharchenko:2013fk}. The offsets are in the sense:  M3: DR1 [Fe/H] $-$ \citet{Harris:1996lr} [Fe/H] = $-$0.15 dex; M13: DR1 [Fe/H] $-$ \citet{Harris:1996lr} [Fe/H] = $-$0.03 dex.

Members and extratidal halo stars of a GC are expected to have a similar [Fe/H] to that of the cluster (for a review see \citealt{Gratton:2012lr}) and most GCs, including M3 \citep{Cohen:2005fk} and M13 (\citealt{Cohen:2005fk} and \citealt{Johnson:2012lr}) do not have a significant [Fe/H] spread. The quoted typical uncertainty in [Fe/H] for LAMOST spectra is 0.1 to 0.2 dex \citep{1674-4527-15-8-1197}, so we initially considered this as a limit for a star to have an [Fe/H] consistent with that of the cluster. However the DR1 Catalog [Fe/H] errors for our candidates are generally considerably higher than this, ranging from 0.23 to 1.16 with a median value of 0.58 dex for M3 and 0.19 to 1.46 with a median value of 0.86 dex for M13. 

\begin{figure*}
\includegraphics[width=168mm]{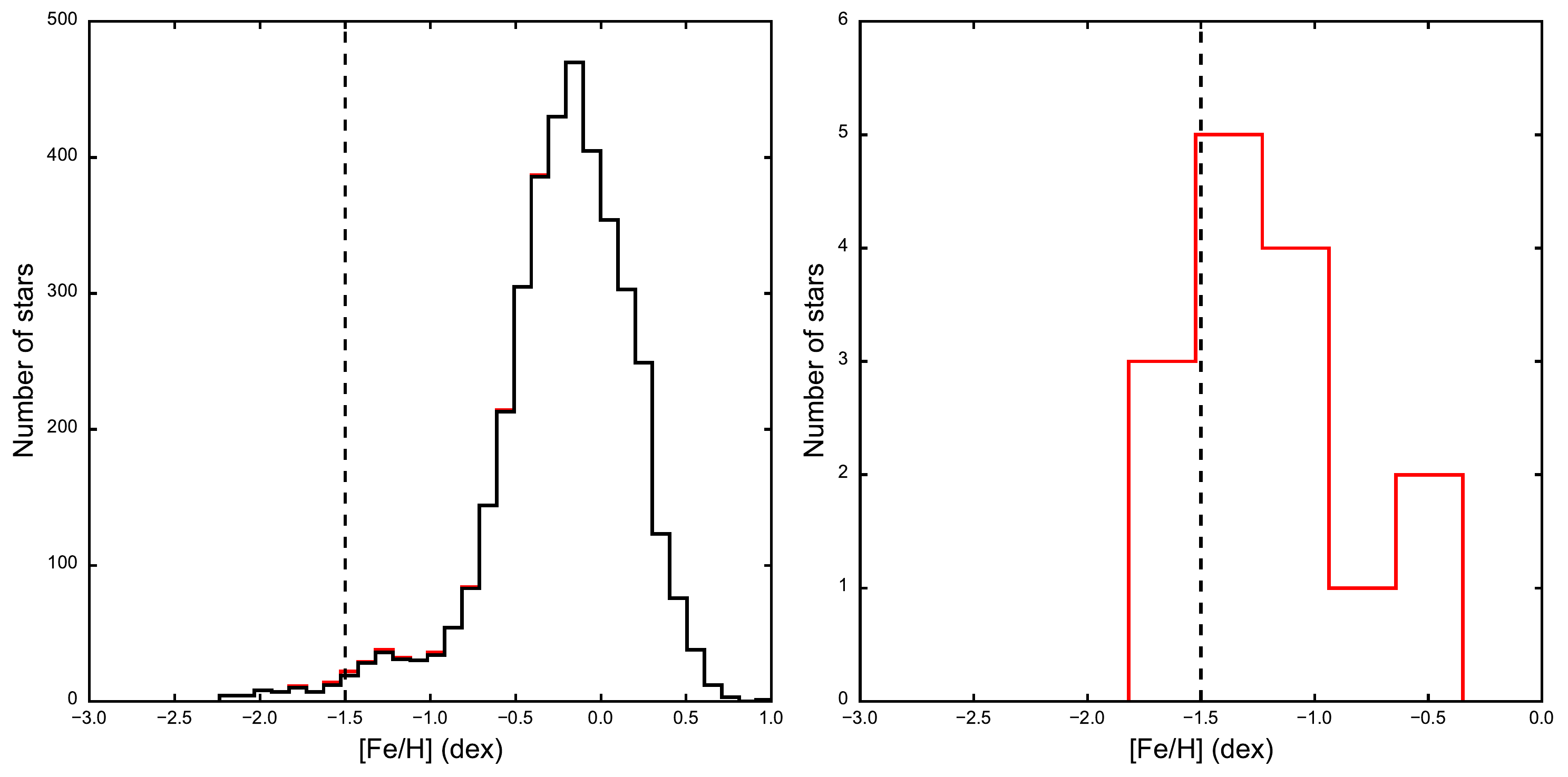}
\caption{Left-hand panel: The black line shows the [Fe/H] distribution of observed stars within a 5\degr\ radius of the M3 central position. Right-hand panel: The red line shows the expanded [Fe/H] distribution of candidate stars. The vertical dashed line indicates the GC [Fe/H].}
\label{NGC5272_M3_FeH_histograms}
\end{figure*}

\begin{figure*}
\includegraphics[width=168mm]{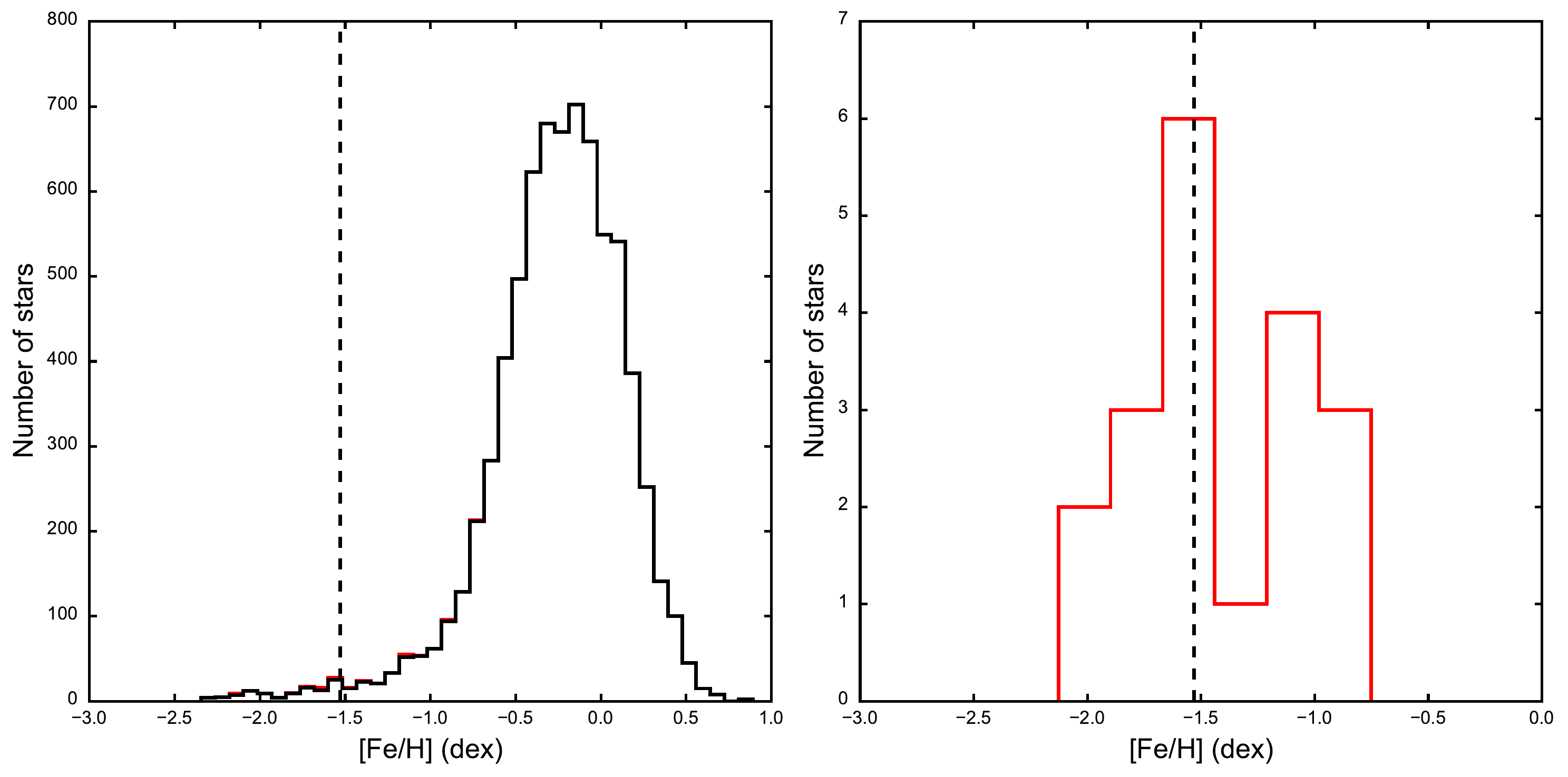}
\caption{Same as Figure \ref{NGC5272_M3_FeH_histograms} for M13.}
\label{NGC6205_M13_FeH_histograms}
\end{figure*}

No studies have shown any overall star-to-star [Fe/H] metallicity variations for M3 \citep{Cohen:2005fk}. Figure \ref{NGC5272_M3_FeH_histograms} shows the [Fe/H] distributions for M3. The left-hand panel shows all the observed stars within a radius of 5\degr\ of the GC central position. The distribution shows a prominent peak consisting of 3879 stars with a mean [Fe/H] $\sim-$0.22 dex that we predominantly identify as Galactic disc or halo field stars (non cluster members). The right-hand panel of Figure \ref{NGC5272_M3_FeH_histograms} is an expanded version centred on the [Fe/H] of the GC showing only the 15 candidate stars. Values range from $-$1.82 to $-$0.35 dex. There are 12 of the 15 candidate stars within the median error ($\pm0.58$ dex) of the GC [Fe/H] of $-$1.50 dex; the bulk of stars are more metal-rich than the cluster [Fe/H]. Values of [Fe/H] in \citet{1674-4527-15-8-1197} for member stars range from $-$2.09 to $-$1.1 dex; they did not eliminate any stars as possible members based on [Fe/H].

M13 is also not known to exhibit any overall [Fe/H] metallicity variations (\citealt{Cohen:2005fk} and \citealt{Johnson:2012lr}). The [Fe/H] distributions for M13 are shown in Figure \ref{NGC6205_M13_FeH_histograms}. The main peak in the left-hand panel consists of 7355 stars with a mean [Fe/H] $\sim-$0.26 dex that we predominantly identify as non cluster members. The right-hand panel of Figure \ref{NGC6205_M13_FeH_histograms} is an expanded version centred on the [Fe/H] of the GC showing only the 19 candidate stars. Values range from $-$2.13 to $-$0.75 dex. All 19 candidate stars  are within the median error ($\pm0.86$ dex) of the GC [Fe/H] of $-$1.53 dex. \citet{1674-4527-15-8-1197} again did not eliminate any stars on the basis of their [Fe/H]; values of member stars of M13 in their study range from $-$1.81 to$-$1.56 dex. 

In view of the large ranges in [Fe/H] and the large errors in [Fe/H] of stars in our samples, we also chose not to eliminate any stars for either cluster based on [Fe/H]. The DR1 Catalog values of [Fe/H] as calculated by LASP do not seem to be helpful in this case. It would be desirable to revisit these limits in future with more precise abundance information.

\section{Spatial distribution}
\label{sd}

\begin{figure*}
\flushleft{\includegraphics[width=168mm]{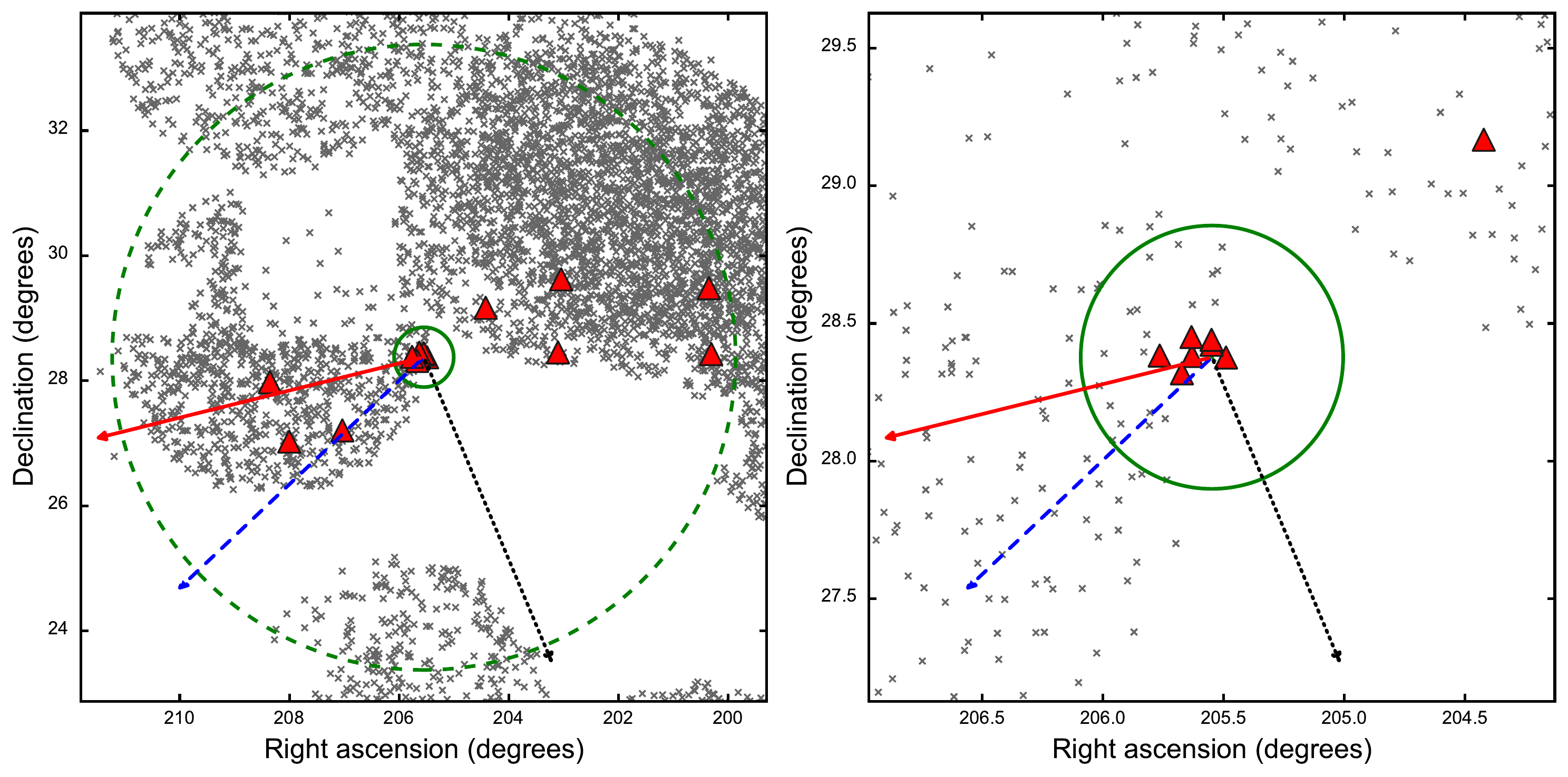}}
\caption{Left-hand panel: Spatial distribution of DR1 Catalog stars around M3. Right-hand panel: Expanded version showing the DR1 Catalog stars inside the tidal radius. Grey crosses show non members, red triangles denote candidate cluster member stars and candidate extratidal cluster halo stars. The green cross and solid circle show the centre position and the tidal radius, the dashed circle is at a 5\degr\ radius from the centre position. The black dotted arrow indicates the direction of the cluster proper motion, the blue dashed arrow is the direction to the Galactic centre and the solid red arrow is the direction perpendicular to the Galactic plane.}
\label{NGC5272_M3_RA_Dec}
\end{figure*}

\begin{figure*}
\flushleft{\includegraphics[width=168mm]{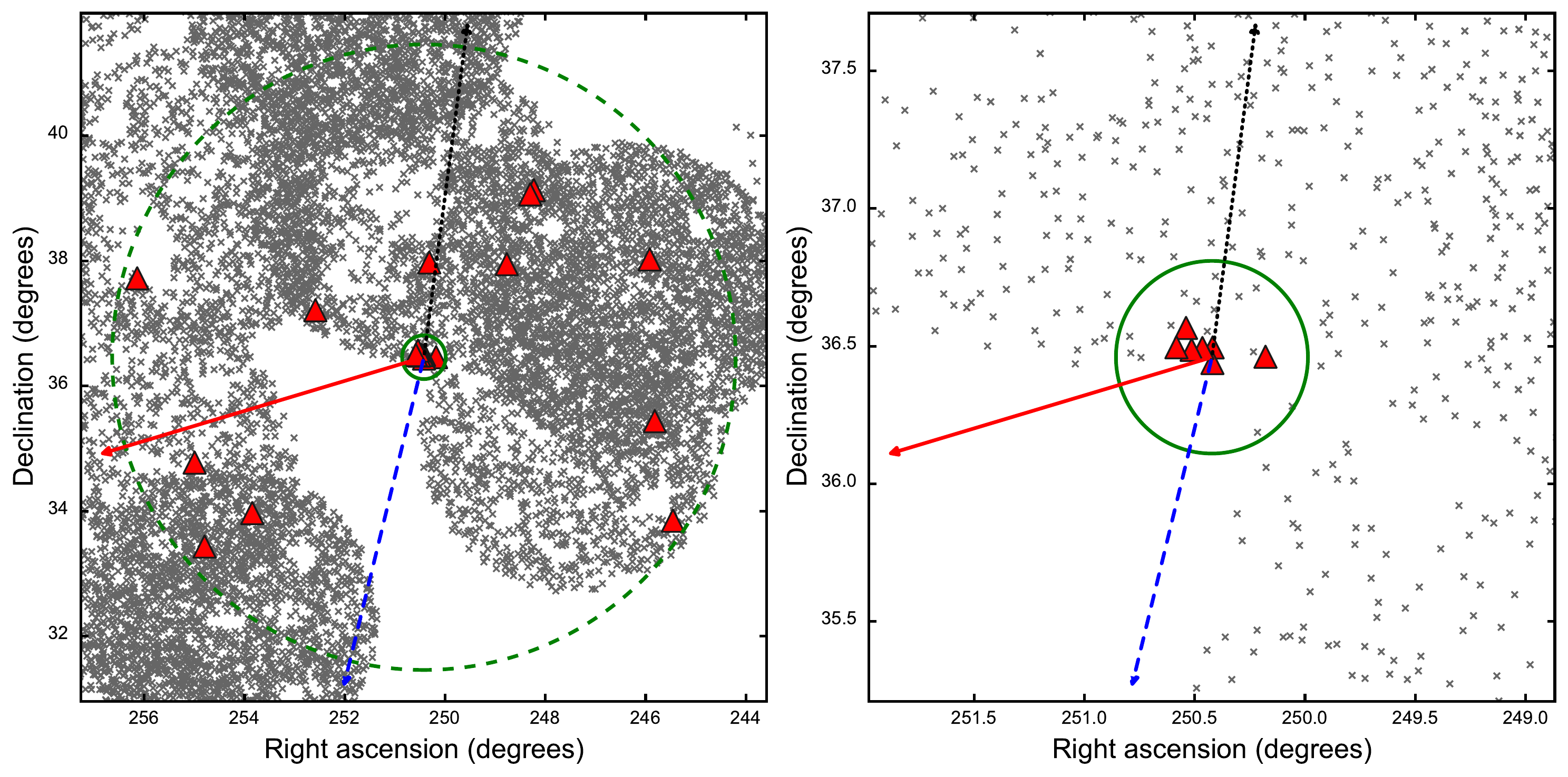}}
\caption{Same as Figure \ref{NGC5272_M3_RA_Dec} for M13.}
\label{NGC6205_M13_RA_Dec}
\end{figure*}

Figure \ref{NGC5272_M3_RA_Dec} shows the spatial (RA - Dec) distribution of observed and candidate stars for M3. Of the 15 final candidate stars, eight are outside the adopted tidal radius of 28.7 arcmin at distances from the GC centre ranging from $\sim$2.6 to $\sim$9.8 times the tidal radius.  Seven candidate stars are inside the tidal radius and we recovered seven of the nine unique stars identified as cluster members in \citet{1674-4527-15-8-1197}. There are a further two stars identified in that paper (based on DR2) that are not in the DR1 Catalog. The eight stars outside the adopted tidal radius are identified as candidate extratidal cluster halo stars.  The stars appear to form a broadly symmetrical linear feature on both sides of the cluster, and it is possible that this is an indication of a tidal tail. However the alignment is roughly perpendicular to the direction of the cluster proper motion and the tails should be extended along the cluster orbit. It also appears to be somewhat aligned with the coverage of the fields, although the North-East quadrant looks fairly uniform. More complete sky coverage as the LAMOST survey continues should resolve this uncertainty.

The spatial distribution of stars for M13 is shown in Figure \ref{NGC6205_M13_RA_Dec}. There are 12 of the 19 candidate stars outside the adopted tidal radius of 21.0 arcmin. For these stars the distances range from $\sim$4.3 to $\sim$13.8 times the tidal radius. Seven of the candidate stars are inside the tidal radius and we recover all four of the stars identified as cluster members in \citet{1674-4527-15-8-1197}. The 12 stars outside the adopted tidal radius are identified as candidate extratidal cluster halo stars.

\section{Final member list and discussion}
\label{discussion}

There are a number of parameters that can be used to select candidate members of a GC from a field sample of stars. Here we had position, radial velocities, stellar atmospheric parameters ($T_{eff}$, log$g$ and [Fe/H]), photometry and proper motions. We used $V_r$ as the primary discriminant, but we have constructed a composite filter, based on all available information, that gives better member/non-member discrimination than any single parameter.

Our first step was to identify the characteristics of GCs that were likely to have member stars in the DR1 Catalog.  Spatially they had to be within the survey area of LAMOST (declination $-$10\degr\ to +90\degr) and we then chose to use GCs that had relatively high heliocentric radial velocities ($|V_r| >100$ \kms). Stars from these GCs should have significantly different $V_r$ to field stars in the same region of sky, and we made an initial search for stars with $V_r$ within \error20~\kms~of the respective GC $V_r$. To search for possible cluster halo stars or tidal tails well outside the GC tidal radius we chose a wide search area encompassing a radius of 5\degr\ from the GC central position. After eliminating GCs that only had small numbers of candidate stars, or where the differentiation between the GC $V_r$ and field stars was not large or the observed stars were too bright to be cluster members, we were left with M3 and M13 as likely candidate GCs to search for extratidal stars.

The first criterion for selection as a cluster member was $V_r$. The range for selection was based on the standard deviation $\sigma$ of Gaussian fits to probability distribution functions of $V_r$. M3 has 50 unique candidate members in LAMOST DR1 that are within a radius of 5\degr\ of the GC central position and that have a $V_r$ within \error$2~\sigma$ of the GC $V_r$. Based on $V$ versus $V-K$ and log($T_{eff}$) versus log$g$ limits, we eliminated 31 of these candidates. For the 19 remaining stars, there are seven stars inside and 12 stars outside the tidal radius. For the 12 stars outside the tidal radius, four have high proper motions relative to the GC, and we accepted the remaining eight stars with proper motions within 10~mas~yr$^{-1}$ of the GC proper motion. We discounted the DR1 Catalog [Fe/H] as a selection criterion.  We identify the remaining eight stars as our final cluster halo member sample. Spatially these stars range from $\sim$2.6 to $\sim$10.2 times the tidal radius. For the seven stars inside the tidal radius we did not use proper motions so these stars are our final cluster members. We recovered all seven of the stars identified as cluster members in \citet{1674-4527-15-8-1197} that are in DR1. All the candidate member and extratidal halo stars are listed in Table~\ref{M3_table}; the final column gives their status as: m = candidate cluster member star, h = candidate extratidal cluster halo star. Altogether we eliminated 35 stars out of the 50 that were selected on the basis of $V_r$ alone, so this is not inconsistent with the predicted $\sim$18 stars unrelated to the cluster that would be in the same region of sky and within \error$2~\sigma$ of the GC $V_r$ from the Milky Way model.

We identified 61 candidate members of M13 in LAMOST DR1 from radial velocities alone. We eliminated 29 of these candidates based on $V$ versus $V-K$ and log($T_{eff}$) versus log$g$ limits and 13 based on proper motion limits. There are 12 of the remaining 19 stars are outside the tidal radius at distances ranging from $\sim$3.4 to $\sim$13.8 times the tidal radius and these stars make up our final candidate cluster extratidal halo sample. The seven stars inside the tidal radius make up our final cluster member sample. We recovered all four of the stars identified as cluster members in \citet{1674-4527-15-8-1197}.  We eliminated 42 stars out of the 61 stars that were selected on the basis of $V_r$ alone and, again, this is reasonably consistent with our Milky Way model predicted sample contamination of $\sim$28 stars.  All the candidate member and extratidal halo stars are listed in Table~\ref{M13_table}, the final column gives their status.
 
\onecolumn
\begin{sidewaystable}
\setlength\tabcolsep{2pt}
\centering
\caption{Candidate cluster members and cluster halo stars of M3 (NGC~5272)}
\label{M3_table}
\begin{tabular}{l l c c c c c c c c c c c c c}
\hline
obsid & ID & RA & Dec & V & K$_s$ & $V_r$ & $T_{eff}$ & log$g$ & [Fe/H] & $\mu_{\alpha}cos(\delta)$ & $\mu_{\delta}$ & R & $P_c$ & Status\\
 & & (degrees) & (degrees) & (mag) & (mag) & (\kms) & (K) & & & (mas~yr$^{-1}$) & (mas~yr$^{-1}$) & (arcmin) & & \\ [0.5ex]
\hline
51313025\textsuperscript{1}     & 2MASS J13211409+2825469  &  200.308734  &  28.42971  &  16.18  &  13.97  &  $-$136.7  &  5012  &  2.2  &  $-$1.54  &  $-$3.2  &  0.6  &  276.5  &  0.79  &  h  \\
145202042 & 2MASS J13212483+2928571  &  200.353534  &  29.48254  &  13.79  &  11.34  &  $-$117.7  &  4604  &  1.3  &  $-$1.00  &  $-$2.1  &  $-$5.5  &  280.7  &  0.02  &  h  \\
145207028 & 2MASS J13321118+2937216  &  203.046613  &  29.62271  &  14.74  &  12.67  &  $-$121.0  &  5011  &  2.2  &  $-$0.60  &  $-$3.3  &  $-$3.6  &  151.1  &  0.06  &  h  \\
48701211  & 2MASS J13322499+2827163  &  203.104159  &  28.45454  &  14.93  &  12.10  &  $-$131.6  &  4586  &  1.8  &  $-$0.81  &  1.0  &  $-$2.5  &  129.1  &  0.47  &  h  \\
48707124  & 2MASS J13374132+2910037  &  204.422183  &  29.16770  &  14.77  &  12.82  &  $-$117.2  &  5081  &  3.4  &  $-$0.35  &  $-$1.5  &  $-$4.3  &  75.9  &  0.02  &  h  \\
110310139 & Cl* NGC 5272 SK 533      &  205.490031  &  28.37642  &  14.14  &  11.99  &  $-$153.2  &  4803  &  1.7  &  $-$1.49  &  $-$6.5  &  9.1  &  3.1  &  0.55  &  m  \\
110310150 & Cl* NGC 5272 SK 326      &  205.549885  &  28.42252  &  14.64  &  12.11  &  $-$148.5  &  4769  &  1.8  &  $-$1.38  &  $-$1.7  &  2.9  &  2.7  &  0.85  &  m  \\
110210143 & Cl* NGC 5272 SK 675      &  205.550218  &  28.44069  &  12.63  &  9.20  &  $-$147.6  &  4199  &  0.7  &  $-$1.17  &  $-$6.1  &  $-$5.3  &  3.8  &  0.89  &  m  \\
110310142 & Cl* NGC 5272 S 1104      &  205.630851  &  28.38177  &  13.70  &  11.08  &  $-$152.6  &  4419  &  1.0  &  $-$1.43  &  $-$4.6  &  3.5  &  4.4  &  0.59  &  m  \\
110310134 & Cl* NGC 5272 SK 68       &  205.633223  &  28.45072  &  13.98  &  11.42  &  $-$152.8  &  4480  &  1.1  &  $-$1.49  &  0.5  &  $-$3.6  &  6.3  &  0.58  &  m  \\
110210146 & Cl* NGC 5272 SK 26       &  205.672917  &  28.31891  &  12.66  &  9.12  &  $-$155.3  &  4099  &  0.7  &  $-$0.95  &  0.9  &  $-$2.2  &  7.4  &  0.42  &  m  \\
110210131 & 2MASS J13430354+2823003  &  205.764840  &  28.38346  &  12.83  &  9.60  &  $-$148.6  &  4215  &  0.7  &  $-$1.22  &  $-$0.6  &  1.1  &  11.4  &  0.84  &  m  \\
110202247 & 2MASS J13480854+2712278  &  207.035611  &  27.20771  &  14.10  &  11.61  &  $-$141.5  &  4635  &  1.3  &  $-$1.29  &  $-$3.7  &  $-$11.9  &  105.6  &  0.99  &  h  \\
110201055\textsuperscript{1} & 2MASS J13520088+2701435  &  208.003680  &  27.02878  &  14.03  &  11.48  &  $-$153.0  &  4686  &  1.4  &  $-$1.82  &  $-$4.7  &  $-$1.0  &  153.5  &  0.56  &  h  \\
110305113 & 2MASS J13532551+2758396  &  208.356275  &  27.97771  &  15.44  &  13.17  &  $-$149.2  &  4866  &  2.0  &  $-$1.57  &  6.2  &  1.0  &  150.4  &  0.80  &  h  \\
\hline
\multicolumn{14}{l}{\footnotesize{R = radial distance from cluster centre.}} \\
\multicolumn{14}{l}{\footnotesize{$P_c$ = membership probability based on $V_r$.}} \\
\multicolumn{14}{l}{\footnotesize{Status: m = candidate  cluster member star, h = candidate extratidal cluster halo star.}} \\
\multicolumn{14}{l}{\footnotesize{Identifications are from \citet{Sandage:1953lr}, \citet{Sandage:1982lr} or 2MASS.}} \\
\multicolumn{13}{l}{\textsuperscript{1}\footnotesize{Duplicate of 110301055}} 
\end{tabular}
\end{sidewaystable}

\begin{sidewaystable}
\setlength\tabcolsep{2pt}
\centering
\caption{Candidate cluster members and cluster halo stars of M13 (NGC~6205)}
\label{M13_table}
\begin{tabular}{l l c c c c c c c c c c c c c}
\hline
obsid & ID & RA & Dec & V & K$_s$ & $V_r$ & $T_{eff}$ & log$g$ & [Fe/H] & $\mu_{\alpha}cos(\delta)$ & $\mu_{\delta}$ & R & $P_c$ & Status\\
 & & (degrees) & (degrees) & (mag) & (mag) & (\kms) & (K) & & & (mas~yr$^{-1}$) & (mas~yr$^{-1}$) & (arcmin) & & \\ [0.5ex]
\hline
154202032  & 2MASS J16214912+3350517    &  245.454661  &  33.84772  &  12.95  &  9.87  &  $-$262.2  &  4233  &  0.4  &  $-$2.13  &  1.2  &  $-$3.8  &  289.7  &  0.42  &  h  \\
50501031   & 2MASS J16231617+3526204    &  245.817393  &  35.43900  &  $-$  &  13.51  &  $-$235.6  &  5200  &  2.4  &  $-$2.10  &  ...  &  ...  &  231.9  &  1.00  &  h  \\
50515195   & 2MASS J16234061+3801295    &  245.919214  &  38.02485  &  15.88  &  13.60  &  $-$226.6  &  5081  &  2.9  &  $-$1.06  &  $-$6.7  &  3.6  &  234.6  &  0.90  &  h  \\
50512192   & 2MASS J16325546+3907555    &  248.231107  &  39.13208  &  15.42  &  13.27  &  $-$267.9  &  4943  &  2.2  &  $-$1.49  &  $-$5.1  &  $-$1.9  &  191.0  &  0.28  &  h  \\
46412198\textsuperscript{1}   & 2MASS J16331239+3903510    &  248.301662  &  39.06417  &  15.09  &  12.82  &  $-$230.2  &  4737  &  1.6  &  $-$1.77  &  $-$5.5  &  3.9  &  185.8  &  0.96  &  h  \\
50513129   & 2MASS J16350451+3756531    &  248.768818  &  37.94810  &  15.82  &  13.61  &  $-$216.8  &  4883  &  3.3  &  $-$0.93  &  $-$9.9  &  1.5  &  119.2  &  0.63  &  h  \\
51801233   & Cl* NGC 6205 SANDA A1      &  250.179108  &  36.46162  &  13.40  &  10.70  &  $-$241.5  &  4477  &  0.9  &  $-$1.58  &  $-$0.2  &  $-$5.2  &  11.7  &  0.96  &  m  \\
51805041   & 2MASS J16411545+3758307    &  250.314372  &  37.97520  &  15.09  &  12.95  &  $-$221.2  &  4898  &  1.9  &  $-$1.56  &  $-$2.5  &  $-$2.1  &  91.1  &  0.76  &  h  \\
51801212   & 2MASS J16414050+3629442    &  250.418656  &  36.49562  &  ...  &  12.45  &  $-$245.7  &  5013  &  2.4  &  $-$1.54  &  24.2  &  $-$3.2  &  2.2  &  0.89  &  m  \\
48801219   & Cl* NGC 6205 BARN 123      &  250.419514  &  36.43891  &  ...  &  12.06  &  $-$247.5  &  5404  &  2.6  &  $-$1.37  &  4.9  &  $-$9.9  &  1.3  &  0.84  &  m  \\
48801215   & Cl* NGC 6205 ARP 1054      &  250.464965  &  36.49338  &  ...  &  11.55  &  $-$221.0  &  5115  &  2.0  &  $-$1.18  &  $-$13.1  &  1.0  &  2.9  &  0.76  &  m  \\
51801225   & Cl* NGC 6205 ARP 1067      &  250.513907  &  36.48670  &  14.43  &  12.05  &  $-$244.7  &  4749  &  1.6  &  $-$1.76  &  1.3  &  $-$0.6  &  4.7  &  0.91  &  m  \\
51801203   & 2MASS J16420943+3633591    &  250.539279  &  36.56642  &  15.08  &  12.91  &  $-$236.4  &  4998  &  2.1  &  $-$1.63  &  $-$3.5  &  $-$7.7  &  8.5  &  1.00  &  m  \\
51801215   & Cl* NGC 6205 KAD 669       &  250.582717  &  36.49612  &  14.82  &  12.48  &  $-$242.5  &  4910  &  1.8  &  $-$1.62  &  $-$2.0  &  $-$6.4  &  8.1  &  0.95  &  m  \\
51807066   & 2MASS J16502062+3712378    &  252.585936  &  37.21049  &  14.72  &  12.52  &  $-$260.5  &  5038  &  2.5  &  $-$1.18  &  $-$7.9  &  $-$0.1  &  113.3  &  0.47  &  h  \\
151410043  & 2MASS J16552301+3358099    &  253.845931  &  33.96943  &  13.07  &  10.35  &  $-$228.0  &  4570  &  1.9  &  $-$0.92  &  $-$5.6  &  0.1  &  224.7  &  0.92  &  h  \\
144709105  & 2MASS J16591121+3326004    &  254.796675  &  33.43342  &  14.96  &  12.81  &  $-$261.0  &  5070  &  2.6  &  $-$0.75  &  $-$0.4  &  7.8  &  281.5  &  0.45  &  h  \\
151403167  & 2MASS J16595873+3446331    &  254.994732  &  34.77585  &  14.24  &  12.06  &  $-$214.8  &  4693  &  1.5  &  $-$1.67  &  $-$2.9  &  4.1  &  244.8  &  0.58  &  h  \\
151007113  & 2MASS J17043362+3743195    &  256.140100  &  37.72209  &  14.77  &  12.61  &  $-$259.3  &  5071  &  2.9  &  $-$1.11  &  $-$4.3  &  2.5  &  283.9  &  0.50  &  h  \\
\hline
\multicolumn{14}{l}{\footnotesize{R = radial distance from cluster centre.}} \\
\multicolumn{14}{l}{\footnotesize{$P_c$ = membership probability based on $V_r$.}} \\
\multicolumn{14}{l}{\footnotesize{Status: m = candidate  cluster member star, h = candidate extratidal cluster halo star.}} \\
\multicolumn{14}{l}{\footnotesize{Identifications are from \citet{Sandage:1970fk}, \citet{Barnard:1931qy}, \citet{Arp:1955uq}, \citet{Kadla:1966lr} or 2MASS.}} \\
\multicolumn{14}{l}{\textsuperscript{1}\footnotesize{Duplicate of 505012198}} 
\end{tabular}
\end{sidewaystable}
\twocolumn

If these stars have indeed escaped from their parent clusters, it is of interest to compare the destruction rate that this implies to that of simulations such as \citet{Gnedin:1997lr} and \citet{Moreno:2014qy}. 

To estimate the observed cluster fractional mass loss we used the ratio of the total cluster extratidal halo stars V luminosity to the integrated cluster V luminosity (from the cluster integrated V magnitude). The observed cluster fractional mass losses are 0.0038 and 0.0046 for M3 and M13 respectively. We then adjusted for sample completeness: (i) DR1 is not spatially complete (see Figure \ref{NGC5272_M3_RA_Dec} and Figure \ref{NGC6205_M13_RA_Dec}) over the area of sky we searched, and (ii) DR1 is not photometrically complete to the faintness limit of the RGB for these GCs. We estimated completeness as the ratio of the number of DR1 stars to samples of stars from the UCAC4 \citep{Zacharias:2013lr}  catalog. For both LAMOST and UCAC4 we selected stars outside the GC tidal radius with a faint limit of V $<$ 16 mag. The V magnitudes in UCAC4 are from APASS \citep{Henden:2009qy} so the limit was chosen to match the quoted current completeness of V=16 mag. This gave a DR1 completeness of 0.209 and 0.143 for M3 and M13 respectively. The fractional mass losses were divided by the corresponding completeness to give total fractional mass losses of 0.0167 for M3 and and 0.0196 for M13.

We then estimated the time taken for a star to move outside our 5\degr\ search area. As an estimate, \citet{Kupper:2010fk} Equation (18) gives the relative velocity of escaped stars for clusters in circular orbits in the disc. For M3 this gives a relative velocity of $\pm\sim$6.0\kms\ and for M13 $\pm\sim$7.3\kms. As stars can escape in any direction, the mean relative velocities perpendicular to our line-of-sight (i.e. the proper motions) are $\pm\sim$6.0$\times2/\pi =\ \pm\sim$3.8\kms and $\pm\sim$6.0$\times2/\pi =\ \pm\sim$4.7\kms. The stars would move 5\degr\ from the GC central position in $\sim$228 Myr and $\sim$130 Myr for M3 and M13 respectively at these velocities.

We divided the fractional mass losses by the time taken for the cluster extratidal halo stars to move outside our search area to give cluster destruction rates. For M3 this was $8.04\times10^{-11}\ yr^{-1}$ and for M13 our calculated observed destruction rate was $2.47\times10^{-10}\ yr^{-1}$.

There are several assumptions/estimates in this calculation that should be mentioned: (i) The mass-to-light ratio for the two samples is probably not the same. The integrated cluster luminosity includes a contribution by dwarfs, whereas the observed total luminosity of the cluster extratidal halo stars does not, as they are too faint to detect in LAMOST data. Therefore the observed cluster fractional mass losses are likely underestimated; this would translate to a smaller calculated destruction rate than is actually occurring. (ii) We have made an estimate of DR1 completeness, but this is another possible source of uncertainty. (iii) It is possible that some of the stars we include in our lists of candidates are field stars rather than ex-members of the cluster, so this would lead to an overestimate of cluster destruction rates. (iv) The estimate of the velocities of escaped stars is, as stated, strictly applicable to clusters in circular orbits in the disc; if the actual velocities are lower then the calculated mass loss rates would also be lower. (v) The velocities of escaped stars are constant. Studies have been done of the variation of $V_r$ along the tidal tail of GCs (e.g \citealt{Odenkirchen:2009qy} and \citealt{Kuzma:2015lr} find gradients of 1.0 $\pm$ 0.1 km s$^{-1}$ deg$^{-1}$ for Palomar 5), but without a detailed model for the gravitational potential of the MW's dark matter halo to calculate the orbit of escaped stars, the effect on the destruction rate is difficult to estimate.

Given those limitations in our estimates, our estimated destruction rate for M3 was $\sim$1--2 orders of magnitude larger than the destruction rates calculated by both \citet{Gnedin:1997lr} and \citet{Moreno:2014qy}. For M13 our estimated destruction rate was $\sim$1 order of magnitude larger  than the destruction rates calculated by \citet{Gnedin:1997lr} but $\sim$3 orders of magnitude larger than those calculated by \citet{Moreno:2014qy}. 

Differences between our detection of candidate extratidal stars and previous studies of M3 and M13 are not unexpected. M3 observations in particular have produced ambiguous results with \citet{1538-4357-639-1-L17}, \citet{Jordi:2010lr} and \citet{Carballo-Bello:2014lr} reporting non-detections. However, previous studies are all variations of photometric studies of fairly large numbers of stars. Their analysis used various star counting algorithms and radial density profiles and surface density plots to search for deviations from model profiles and structure or overdensities. These profiles and plots typically did not show deviations or structure much beyond the tidal radius, so the handful of more distant extratidal candidate stars we found would most likely not be detected by these studies. 

Finding significant numbers of stars in the process of escaping from these supposedly relatively stable GCs is intriguing. It furthers the case for the existence of extratidal stars associated with M3 as well as affirming the previous positive results for M13. The discrepancies between our observed destruction rates and the predicted rates call for further investigation. To confirm their status our candidate cluster extratidal halo stars require high resolution spectroscopic observations to match the chemical abundances. The final number confirmed as ex-cluster members can constrain theoretical studies of GC destruction rates as well as the contribution of GCs to the Galaxy's stellar halo. However there are also significant differences in the predicted rates, so the models and simulations are also not yet definitive, and this may also account for some differences.

\section{Conclusions}
\label{conclusions}

We find candidate extratidal stars in wide halos around the globular clusters M3 (NGC~5272) and M13 (NGC~6205) in the LAMOST Data Release 1. If their status is confirmed they support previous studies that both clusters are surrounded by a halo of extratidal stars or exhibit tidal tails. Interestingly, destruction rates corresponding to the observed mass loss are generally significantly higher than theoretical studies would indicate. Large-scale spectroscopic surveys such as LAMOST are ideal for this kind of search, especially when combined with photometric and astrometric data.  More candidate extratidal stars will almost certainly be found for M3 and M13, and possibly other GCs, as the dataset grows, but we support the recommendation of \citep{1674-4527-15-8-1197} to target known photometric members of clusters. High resolution spectroscopic observations of the candidate extratidal cluster halo stars would be valuable in confirming their origin, and hence provide constraints for theoretical studies.

\section*{Acknowledgements}
\label{ack}
We thank Borja Anguiano for his assistance and helpful discussions. The authors also thank the reviewer for their useful comments and suggestions for improvement. SLM and DBZ acknowledge the financial support from the Australian Research Council through grants DE140100598 and FT110100793 respectively. Guoshoujing Telescope (the Large Sky Area Multi-Object Fiber Spectroscopic Telescope LAMOST) is a National Major Scientific Project built by the Chinese Academy of Sciences. Funding for the project has been provided by the National Development and Reform Commission. LAMOST is operated and managed by the National Astronomical Observatories, Chinese Academy of Sciences. This research made use of NASA's Astrophysics Data System; the VizieR catalogue access tool, CDS, Strasbourg, France; \textsc{Astropy}, a community-developed core \textsc{Python} package for Astronomy \citep{Astropy-Collaboration:2013lr}; the \textsc{IPython} package \citep{PER-GRA:2007}; \textsc{SciPy} \citep{jones_scipy_2001}; \textsc{TOPCAT}, an interactive graphical viewer and editor for tabular data \citep{2005ASPC..347...29T}. We acknowledge with thanks the variable star observations from the AAVSO International Database contributed by observers worldwide and used in this research. This publication makes use of data products from the Two Micron All Sky Survey, which is a joint project of the University of Massachusetts and the Infrared Processing and Analysis Center/California Institute of Technology, funded by the National Aeronautics and Space Administration and the National Science Foundation.

{\it Facilities:} \facility{LAMOST}.

\bibliographystyle{apj}
\bibliography{Galactic_archaeology}

\end{document}